\documentclass[a4paper]{article}

\usepackage{INTERSPEECH2022}
\usepackage{amsmath,graphicx}
\usepackage{multirow}
\usepackage[hang,flushmargin]{footmisc}
\usepackage[textfont=it,tableposition=top]{caption}
\usepackage{subcaption}
\usepackage{hyperref}
\hypersetup{colorlinks=true}
\usepackage{lipsum}

\newcommand\blfootnote[1]{%
  \begingroup
  \renewcommand\thefootnote{}\footnote{#1}%
  \addtocounter{footnote}{-1}%
  \endgroup
}

\def \TI                {t}             %
\def \FI                {c}

\def \cleanMel          {\mathbf X}
\def \cleanMelEst       {\widehat{\mathbf X}}
\def \noiseMel          {\mathbf N}
\def \mixMel            {{\mathbf Y}}   %

\def \cleanFeats		{\mathbf f_X}
\def \cleanFeatsEst     {{\mathbf f}_{\widehat{X}}}

\def \encoder           {\mathbf E}
\def \encodert           {\mathbf e_{\TI}}

\def \IRM               {\mathbf M}           %
\def \ERM               {\widehat{\IRM}}   %
\def \scaledERM         {\overline{\IRM}}

\DeclareMathOperator{\sigmoid}{sigmoid}
\DeclareMathOperator{\FCLayer}{FCLayer}
\DeclareMathOperator{\StopGrads}{StopGrads}

\title{Mask scalar prediction for improving robust automatic speech recognition}
\name{Arun Narayanan, James Walker, Sankaran Panchapagesan, Nathan Howard, Yuma Koizumi}
\address{Google LLC, U.S.A.}
\email{\{arunnt, jswalker, panchi, ndhoward, koizumiyuma\}@google.com}

\begin{document}
\ninept

\maketitle
\begin{abstract}
Using neural network based acoustic frontends for improving robustness of streaming automatic speech recognition (ASR) systems is challenging because of the causality constraints and the resulting distortion that the frontend processing introduces in speech. Time-frequency masking based approaches have been shown to work well, but they need additional hyper-parameters to scale the mask to limit speech distortion. Such mask scalars are typically hand-tuned and chosen conservatively. In this work, we present a technique to predict mask scalars using an ASR-based loss in an end-to-end fashion, with minimal increase in the overall model size and complexity. We evaluate the approach on two robust ASR tasks: multichannel enhancement in the presence of speech and non-speech noise, and acoustic echo cancellation (AEC). Results show that the presented algorithm consistently improves word error rate (WER) without the need for any additional tuning over strong baselines that use hand-tuned hyper-parameters: up to 16\%  for multichannel enhancement in noisy conditions, and up to 7\% for AEC.
\end{abstract}

\blfootnote{The authors thank Tom O'Malley, Joe Caroselli, and Alex Park.}
\section{Introduction}
As performance of automatic speech recognition (ASR) systems improved over the years \cite{li2021betterfaster,hsu2021hubert,li2021recent}, the number of applications that use speech as a standard modality of input has increased. With varying uses and high user expectations, an important goal is to ensure that the performance does not deteriorate significantly in harsh acoustic conditions -- something current ASR models still struggle with \cite{barker2018fifthchime}. Such conditions can be the result of significant environmental noise or competing speech. With increasing focus on building large scale, general purpose, multidomain \cite{NarayananMisraSimPundakEtAl18} and multilingual ASR models \cite{pratap2020massively, zhang2021bigssl}, addressing background noise together with variations in domain and language in the same model can lead to additional complexity in training and maintaining ASR models. While data augmentation strategies like SpecAug \cite{park2019specaugment} and multi-condition training \cite{kim2017mtr} help to an extent, it is often advantageous to address environmental noise, device echo and competing speech using dedicated modules. A number of techniques have therefore been proposed in the literature \cite{zhang2018robustasroverview}.

Time-frequency masking based techniques are commonly used to build acoustic frontends for ASR \cite{Narayanan2013IRM}. In a single-channel setting, a time-frequency mask, either in the complex spectral domain \cite{wang2020complex} or the feature domain \cite{Narayanan2013IRM}, is used to estimate clean speech, which is then passed on to the backend ASR model. Alternatively, frontend processing can be done directly in the time-domain \cite{luo2018tasnet,kinoshita2020timedomainforasr}. Typically, such frontends introduce speech distortion. This is partly due to the mismatch in the training criterion, which optimizes some measure of distance between clean and noisy speech, and the final goal, which is ASR. The constraints of streaming, which limits the model's architecture to be unidirectional and causal, makes the task even more challenging. If multiple microphones are available, the time-frequency mask can be used to compute beamforming filters to remove noise. This limits the amount of distortion, and usually yields better results \cite{heymann2016neural}. Nonetheless, for short queries in a streaming setting, the gains are limited even with multiple microphones \cite{heymann2018performance}.
\begin{figure}[t]
  \centering
  \includegraphics[scale=0.45]{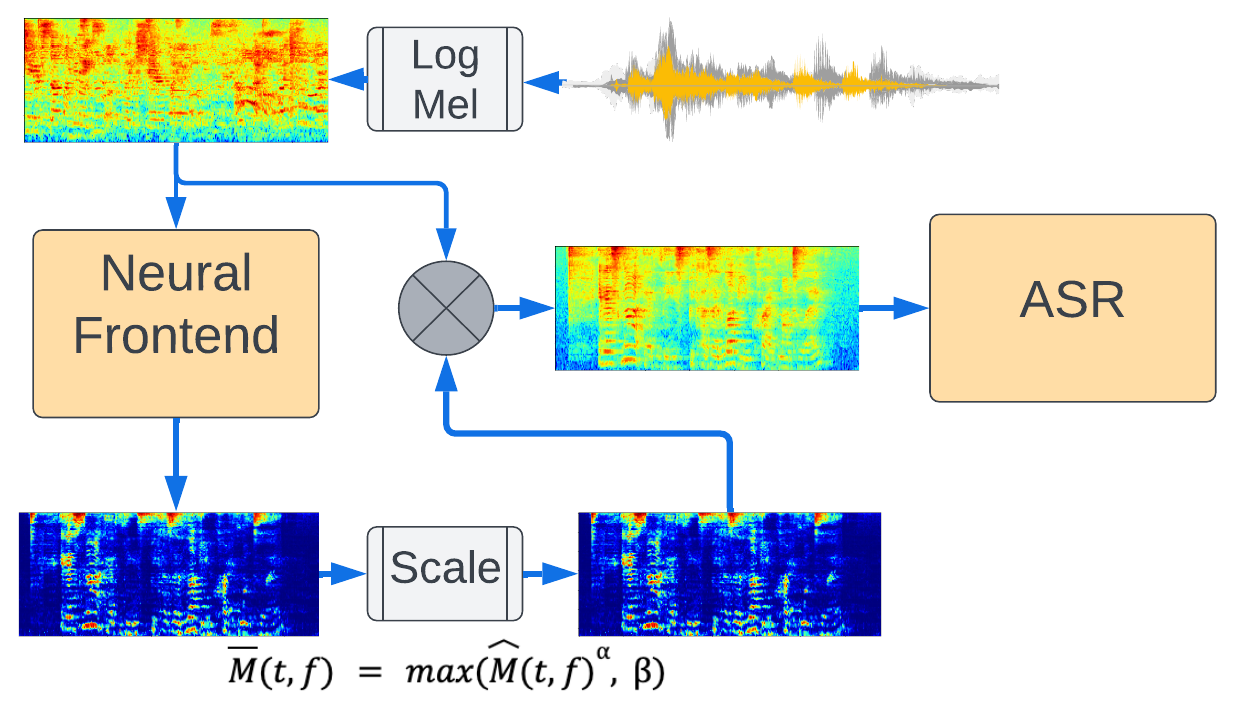}
  \caption{{A block diagram of an acoustic frontend for ASR.}}
   \label{fig:masking}
   \vspace{-0.25in}
\end{figure}

Several techniques have been proposed to reduce speech distortion when using an acoustic frontend for ASR. Menne \textit{et al.} \cite{menne2019investigation} propose estimating the per-frame parameters of a Wiener filter and the noise mask, and joint training with a hybrid ASR model. In \cite{pandey2021dual}, the input to the ASR model is a weighted sum of noisy and enhanced speech features. The weights are computed by a neural network, and jointly trained with ASR. Sato \textit{et al.} \cite{sato2021shouldwe} recommend skipping enhancement in less noisy conditions, which they later extend using a learned weighted sum of noisy and enhanced speech \cite{sato2022learning}. Adding back a scaled version of noisy data is also adopted in \cite{iwamoto2022bad} to address distortions. Koizumi \textit{et al.} \cite{koizumi2021snri} propose using a signal-to-noise ratio improvement (SNRi) target when doing enhancement in the waveform domain. SNRi is predicted by a separate network and optimized using an ASR loss. While most algorithms jointly optimize the the enhancement model and the ASR model \cite{menne2019investigation,pandey2021dual,koizumi2021snri}, there are also approaches that freeze the ASR model \cite{howard2021neural}. We will also use this strategy, since we assume that the ASR model is trained to cover several use cases, and cannot easily be jointly optimized with the enhancement model.

When using a time-frequency mask, one way to limit distortion is to post-process the mask using a mask scalar and a mask floor before using it to enhance the features \cite{Narayanan2014Joint,wang2021task}. The mask scalar exponentially scales the mask, which reduces speech distortion, but retains residual noise. The mask floor limits noise attenuation. One drawback of this approach is that these hyperparameters have to be hand-tuned. Moreover, a single value is chosen in the end, irrespective of the noise condition. In practice, the best value depends on the amount of noise; e.g., in clean conditions, the mask scalar should be close to 0.0 so that the enhanced features are close to the original ``noisy'' features, unlike in noisier conditions. Furthermore, we hypothesize that it may even be preferable to allow the mask scalar to be dynamic, so that its values change from frame-to-frame depending on the level of noise. In this work, therefore, we propose using a \emph{mask scalar net} to predict their optimal values. Similar to \cite{koizumi2021snri}, mask scalar prediction is optimized exclusively using an ASR loss. Our results show that, on two separate robustness tasks -- speech enhancement and acoustic echo cancellation -- using predicted mask scalars provide significant improvements in WER with almost negligible increase model size. 

The rest of the paper is organized as follows. Sec.~\ref{sec:system} provides a detailed description of the proposed models. Experimental settings are described in Sec.~\ref{sec:expr} and results in Sec.~\ref{sec:results}. Finally, conclusions and future work are presented in Sec.~\ref{ref:concl}.

\vspace{-0.1in}
\section{System} \label{sec:system}
\subsection{Masking-based ASR frontend}
\noindent A masking-based acoustic frontend for ASR is shown in Fig.~\ref{fig:masking}. The frontend operates in the log Mel magnitude spectral domain, to match the features used by ASR. 
This not only simplifies joint training with ASR, which is needed for the best performance \cite{Narayanan2014Joint}, but also circumvents the need to estimate phase, in contrast to complex spectral or waveform domain models.

Given a noisy Mel spectrogram, $\mixMel$, the goal is to estimate the clean Mel spectrogram, $\cleanMel$. 
This is done via an estimate of the ideal ratio mask, $\IRM$, which
is the ratio of speech to mixture Mel magnitudes at each time-frequency bin, assuming that speech and noise are uncorrelated: $\IRM(\TI,\FI) = \frac{\cleanMel(\TI,\FI)}{\cleanMel(\TI,\FI) + \noiseMel(\TI,\FI)}.$
Here, $\noiseMel$ is the noise Mel spectrogram, and $\TI$, $\FI$ are time and frequency indices. We assume that $\mixMel \approx \cleanMel + \noiseMel$, so that $\IRM \in [0, 1]$, which simplifies estimation \cite{Narayanan2013IRM}. The mask is estimated using a neural net and post-processed to limit distortions:
\begin{equation}
\scaledERM(\TI,\FI) = \max(\ERM(\TI,\FI)^\alpha, \beta), \label{eq:scale}\\
\end{equation}
$\ERM$ and $\scaledERM$ are the estimated and post-processed masks, respectively.
$\alpha$ and $\beta$ are the hyper-parameters for post-processing: $\alpha \in [0, 1]$ exponentially scales the mask, and $\beta$ floors the mask. 
The estimated clean Mel spectrogram, $\cleanMelEst = \mixMel \odot \scaledERM$, where $\odot$ stands for point-wise multiplication. 
By using a value of $\alpha < 1$ and $\beta > 0$, we can trade-off speech distortion and residual noise.
$\cleanMelEst$ is $\log$ compressed, mean-variance normalized, and, optionally, stacked and subsampled \cite{pundak2016lfr} to create input ASR features. 

We look at two specific frontends in this work: a multichannel enhancement frontend to remove background noise, like a T.V. playing in the background, and an acoustic echo cancellation frontend for removing device echo during playback.
\begin{figure}[t]
     \centering
     \begin{subfigure}[b]{0.22\textwidth}
         \centering
         \includegraphics[width=\textwidth]{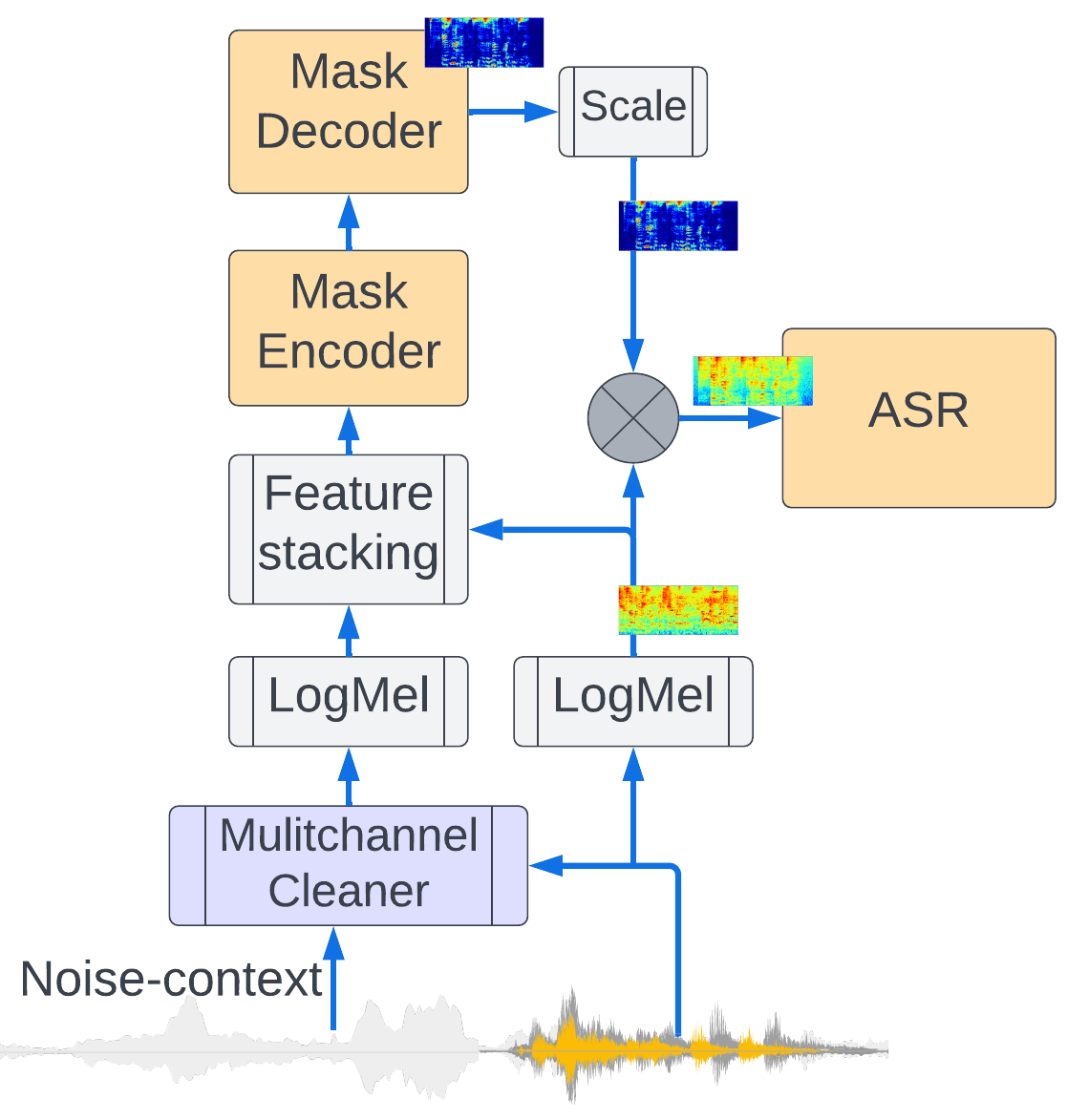}
         \caption{}
         \label{fig:enhancement}
     \end{subfigure}
     \hfill
     \begin{subfigure}[b]{0.22\textwidth}
         \centering
         \includegraphics[width=\textwidth]{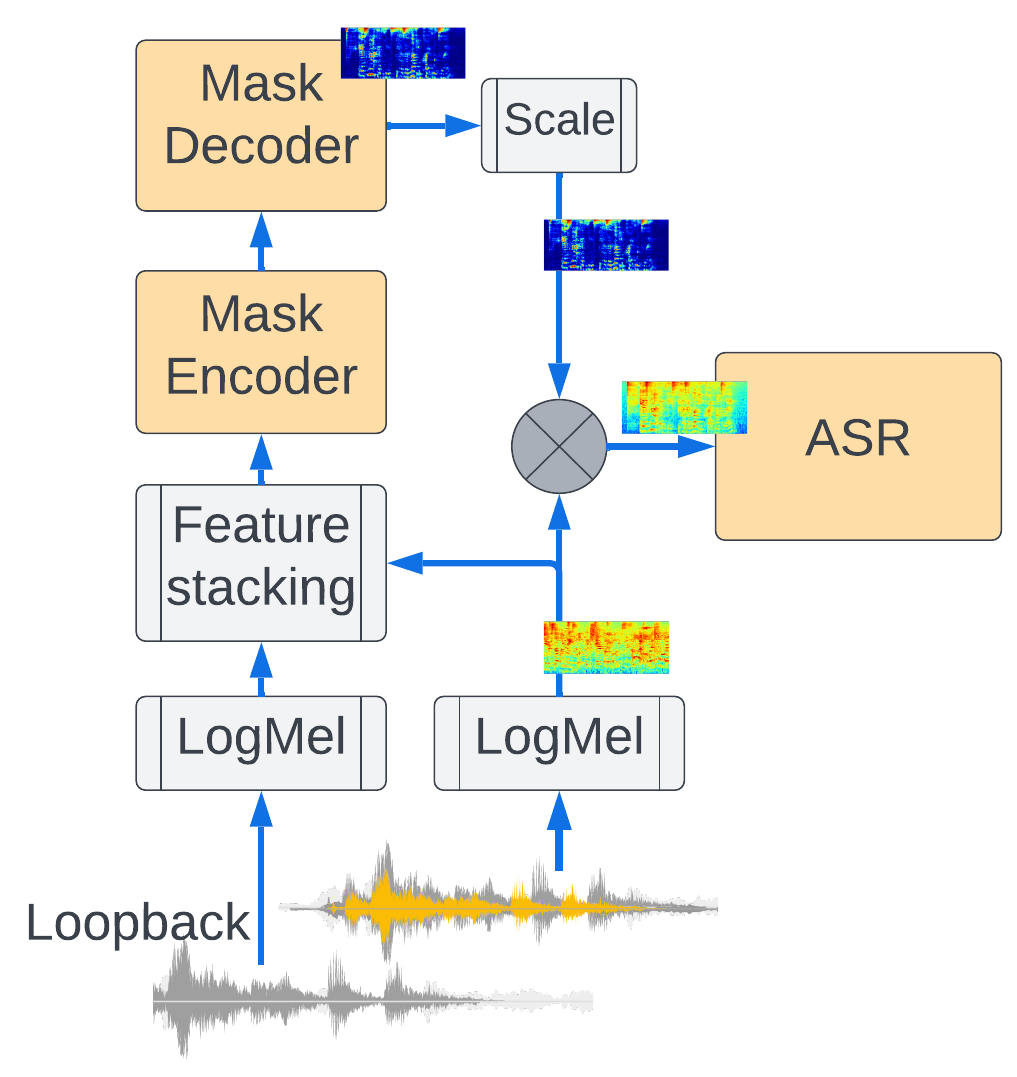}
         \caption{}
         \label{fig:aec}
     \end{subfigure}
        \caption{(a) Cleanformer based enhancement frontend for ASR; (b) Neural acoustic echo cancellation frontend for ASR.}
        \label{fig:enhancement_aec}
   \vspace{-0.2in}
\end{figure}

\subsubsection{Speech Enhancement frontend}\label{sec:cleanformer}
The multichannel enhancement frontend we consider in this work is the recently proposed \emph{Cleanformer} \cite{caroselli2022cleanformer}, as shown in Fig.~\ref{fig:enhancement}. Cleanformer uses Multichannel Cleaner (McCleaner) \cite{huang2019hotwordcleaner}, which estimates an enhanced waveform from multichannel noisy audio and the noise context (a short segment of noise preceding the input to be recognized). 
McCleaner achieves excellent noise reduction, but also introduces distortion, making it less suited for ASR \cite{caroselli2022cleanformer}.
As input, Cleanformer concatenates the log Mel features computed from the output of McCleaner and one of the unprocessed microphone channels. It estimates $\IRM$ using a neural net, which consists of a Conformer encoder \cite{gulati2020conformer} and a fully connected layer ($\FCLayer$) as the mask decoder. The $\FCLayer$ uses a sigmoid activation function: 
\begin{equation}
\ERM(\TI,\cdot) = \sigmoid(\FCLayer(\encodert; \theta_{\IRM})).
\end{equation}
$\encodert$ is the output of the encoder $\TI$ and $\theta_{\IRM}$ the parameters of the $\FCLayer$. Eq.~\ref{eq:scale} is then used for post-processing $\ERM$.
The model is trained using a combination of direct mask loss, $\ell_{\IRM}$, and an ASR loss, $\ell_{ASR}$, \cite{howard2021neural}, using a Conformer-based ASR model \cite{li2021betterfaster}:
\begin{align}
\ell_{\IRM} &= 
\lVert
\IRM - \ERM
\rVert_{1}
+
\lVert
\IRM - \ERM
\rVert_{2} ^2
\nonumber \\
\ell_{ASR}  &= \lVert \encoder_{ASR}(\cleanFeats) - \encoder_{ASR}(\cleanFeatsEst)\rVert_{2} ^2,  \nonumber \\
\ell &= \ell_{\IRM} + \lambda_{ASR} \times \ell_{ASR}. \label{eq:total_loss}
\end{align}
Here, $\lVert {\mathbf A} \rVert_{p} = (\sum_{i,j} \lvert {\mathbf A}(i,j) \rvert ^p)^{\frac{1}{p}}$ is the entry-wise matrix $p$-norm and $\ell$ is the total loss.
$\ell_{ASR}$ is the squared-error between the output of a pre-trained ASR encoder when using clean features, $\cleanFeats$, and the estimate of the clean features, $\cleanFeatsEst$, computed from $\cleanMelEst$. $\lambda_{ASR}$ weights $\ell_{ASR}$. Note that $\ell_{ASR}$ only affects the parameters of the enhancement model; ASR model parameters are kept frozen during training.

\vspace{-0.1in}
\subsubsection{Acoustic Echo Cancellation}
AEC follows an architecture that is very similar to the Cleanformer, as shown in Fig.~\ref{fig:aec}. The main difference is the inputs to the AEC frontend: log Mel features computed on the loopback signal concatenated with those computed on the microphone input. For AEC, we assume that the input is single channel; the goal is to remove the device echo from the microphone input. Other than the inputs, the rest of the architecture closely follows the enhancement frontend; a Conformer encoder followed by an $\FCLayer$ to compute $\IRM$. As before, $\ERM$ is post-processed to $\scaledERM$, and used for estimating clean features for ASR. The model is also trained on the combination of $\ell_{\IRM}$ and $\ell_{ASR}$.

\subsection{Mask-scalar prediction}\label{sec:mask-scalar}
\begin{figure}[ht]
  \centering
  \includegraphics[scale=0.45]{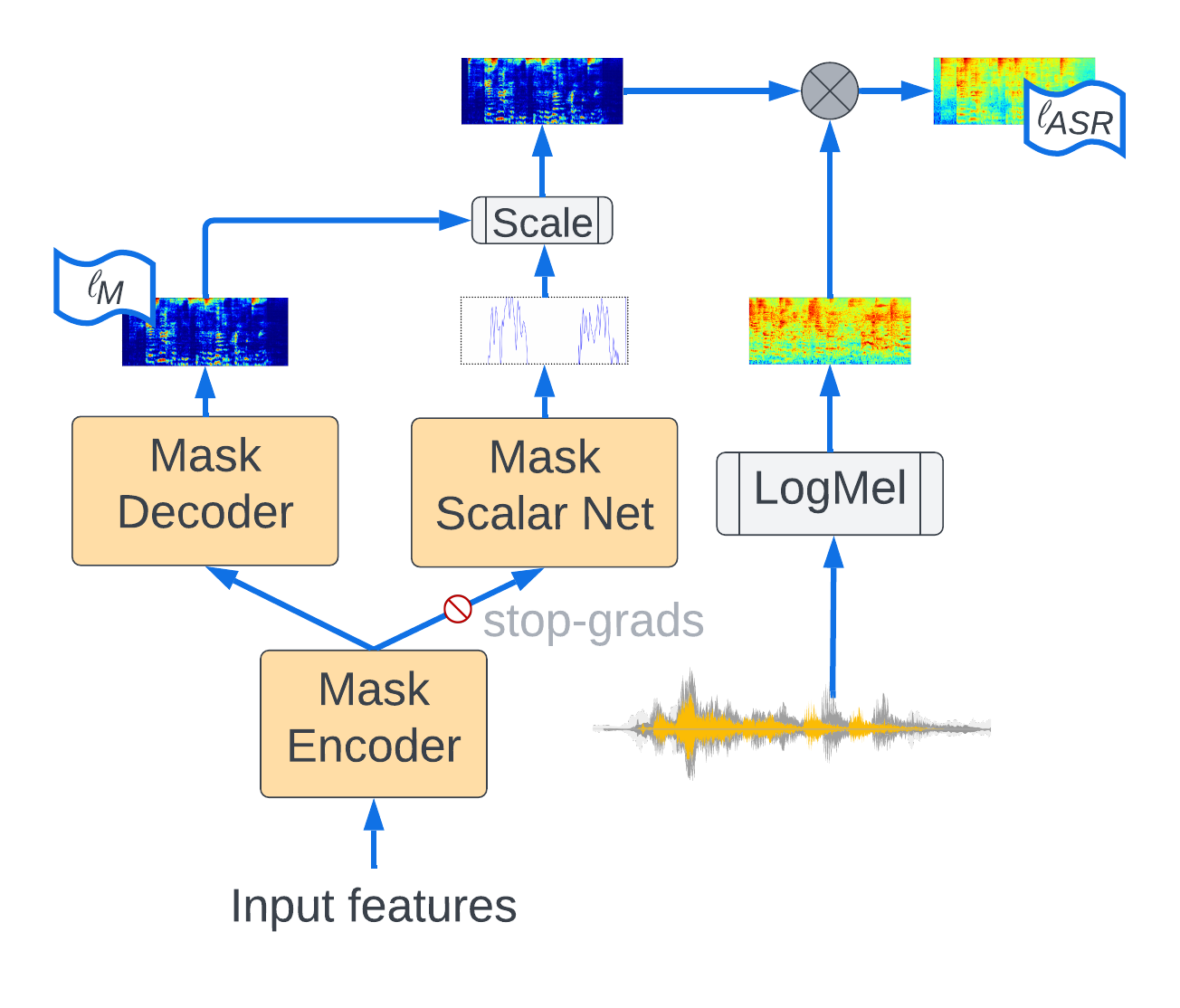}
  \caption{{Mask scalar prediction.}}
   \label{fig:mask-scalar}
   \vspace{-0.2in}
\end{figure}
\noindent Both acoustic frontends use hyperparameters $\alpha$ and $\beta$ to limit speech distortion. The optimal values for $\alpha$ and $\beta$ depend on the input. In clean conditions, $\alpha = 0$ ensures that the input features are unchanged. But in noisy conditions, a larger value of $\alpha$ works better, as we show in Sec.~\ref{sec:results-alpha}. Since the best value is input dependent, we propose learning it directly from the data.

In this work, we predict $\alpha(\TI)$, a per-frame exponential scalar for the mask. The architecture we use is shown in Fig.~\ref{fig:mask-scalar}. Compared to the models described in the previous sections, the model has an additional Mask Scalar Net, which is an $\FCLayer$ that maps the encoder output, $\encodert$, to $\alpha(\TI)$:
\begin{equation}
\alpha(\TI) = \sigmoid(\FCLayer(\StopGrads(\encodert); \theta_{\alpha})).
\end{equation}
$\theta_{\alpha}$ are the parameters of this $\FCLayer$. Importantly, $\theta_{\alpha}$ is learned exclusively using $\ell_{ASR}$. The mask decoder and mask scalar nets share the encoder, which limits the number of additional parameters. However, gradients are not back propagated from the mask scalar net to the encoder ($\StopGrads$). 
This ensures that mask scalar prediction is driven entirely based on ASR performance without being influenced by $\ell_{\IRM}$. Given $\alpha(\TI)$, the post-processed mask is computed as:
\begin{equation}
\scaledERM(\TI,\cdot) = \max(\ERM(\TI,\cdot)^{\alpha(\TI)}, \beta).
\end{equation}
$\scaledERM$ is finally used to obtain the inputs to the ASR model.

\section{Experimental Settings} \label{sec:expr}
\subsection{Data}
The target speech data for training both frontends is created using Librispeech \cite{panayotov2015librispeech}, Librivox \cite{kearns2014librivox} and internal, vendor-collected datasets. 
The training set has 50.8k hours of speech.

\subsubsection{Speech Enhancement}
The training data is generated using a room simulator. Noise is sampled from internally collected noise-only data that simulate conditions like cafe, kitchen, and cars, as well as from Getty\footnote{\url{https://www.gettyimages.com/about-music}} and YouTube Audio Library\footnote{\url{https://youtube.com/audiolibrary}}. The signal to noise ratio (SNR) is sampled from the range -10~dB to 30~dB. Simulated rooms have reverberation times (T60) from 0~msec to 900~msec. The room impulse responses (RIR) correspond to a 3 microphone array, spaced along an equilateral triangle, 66 millimeters apart. We also use multi-talker training sets, using Librivox and vendor-collected utterances as background speech, and at SNRs (defined as target to background-speech) between 1~dB and 10~dB. Multiple copies of the training data are generated, using randomly selected room configurations and SNRs for each target utterance. SpecAug \cite{park2019specaugment} is also used during training.

The test sets consist of: 1)~\emph{Fully simulated} sets based on the test-clean subset of Librispeech; we create a reverberation-only set (``Reverb''), and sets at  -5~dB and 0~dB using environmental noise or speech background.
2)~\emph{Rerecorded v1}, constructed by recording speech and noise separately in a room using a 3-mic array, and mixed at 0~dB and 6~dB SNR; we use pink noise, which is different from the noise types seen in training. These sets have $\sim$13k utterances each. 3)~\emph{Rerecorded v2} records target speech and noise, played simultaneously, using a 5-mic array unseen during training; this set is recorded in 3 conditions: With no additional noise other than the ambient noise in the room (Reverb), with a movie playing in the background (``Movie'') and with babble noise played out from a different location in the room (``Babble''). These sets have 500 utterances each. 

Each utterance in the training and test sets have $\sim$6 seconds of noise context that is used by McCleaner. In multi-talker conditions, the noise context helps identify background speech, since only the interfering speaker is active in the noise context.

\subsubsection{Acoustic Echo Cancellation}
For AEC, we construct training data using synthetic and real echoes \cite{howard2021neural}. To simulate synthetic echoes, both target and loopback signals are convolved with synthetic RIRs, with the hypothetical loudspeaker close to the microphone. The synthetic datasets are created using Librispeech and Librivox as the source data; the same sets are also used as loopback along with waveforms from Getty and YouTube Audio Library. The signal-to-echo ratio (SER) is set to be between -20~dB and 5~dB. To create datasets with real echoes, we rerecorded, at various loudness levels, Librispeech and an internal dataset that was collected for training text-to-speech (TTS) models using Google Home devices. The recorded echoes are added to reverberant target speech at SERs ranging from -20~dB to 5~dB.

Real echoes, recorded using a held-out subset of the utterances from the internal TTS-based dataset, are mixed with the test-clean subset of Librispeech to create test sets. Test sets are constructed at SNRs from -10~dB to 5~dB in 5~dB increments.
\vspace{-0.1in}
\subsection{Architecture}
We use 128-dimensional log Mel features, stacked across 4 contiguous frames as input to the frontends and ASR. The window size is 32~msec, at 10~msec hops. The features are also subsampled by a factor of 3.
All models use the ideal ratio mask corresponding to the reverberant speech as the target.%

The encoder for all our frontends consists of a 4-layer conformer model. Each layer has 256 units; a convolutional block with kernel size 15, and 1024-dimensional feed-forward nets. We use causal masked attention with a left context of 31 frames. The $\FCLayer$ in the mask decoder projects the 256-dimensional encoder output to a 512-dimensional mask. The mask scalar net is a $256\times1$ dimensional $\FCLayer$ that projects the encoder output to a 1-dimensional scalar. In total, the frontends have $\sim$6.5 million parameters. 

$\lambda_{ASR}$ in Eq.~\ref{eq:total_loss} is increased from $0.0$ to $100.0$ linearly from 20k steps to 200k steps, and kept fixed after that. When using the mask scalar net, we use a fixed value of $\alpha = 0.5$ till 200k steps, after which the model optimizes it using the ASR loss. Training $\alpha$ from scratch resulted in divergence. $\theta_{\alpha}$ is initialized by sampling from a Gaussian distribution with a standard deviation of $0.01$. Together with the sigmoid activation used by the FCLayer, this ensures that the initial values for the predicted $\alpha(\TI)$ during training after 200k steps are close to $0.5$ to avoid any sudden jumps that can cause training instability.

The ASR model used for evaluation is an LSTM-based multidomain recurrent neural net transducer \cite{sainath2020streaming}, trained on $\sim$400k hours of English speech, covering domains like near-field and far-field VoiceSearch, YouTube, and Telephony. The utterances for VoiceSearch and Telephony are anonymized and hand-transcribed. We augment training data using SpecAug \cite{park2019specaugment} or simulated noise \cite{kim2017mtr}, making the model robust to moderate noise levels. 
We use the Lingvo toolkit for training \cite{ShenNguyenWuChenEtAl19}.
\section{Results} \label{sec:results}

\begin{table*}[tb]
  \centering
  \caption{Enhancement performance, in terms of ASR WER, using various models. Cleanformer uses $\alpha=0.5, \beta=0.01$. MSP stands for mask scalar prediction. E1, E2 use the model in Fig.~\ref{fig:mask-scalar}; E1 without $\StopGrads$, E2 with $\StopGrads$. E1 and E2 use $\beta=0.01$.}
  \label{tab:enh}
  \begin{tabular}{l|ccccc|cc|ccc}
    \hline
    \multirow{3}{*}{\textbf{Model}} & \multicolumn{5}{c}{\textbf{Librispeech}} & \multicolumn{2}{c}{\textbf{Rerecorded v1}} & \multicolumn{3}{c}{\textbf{Rerecorded v2}} \\
    {} & {} & \multicolumn{2}{c}{\textbf{Environment}} & \multicolumn{2}{c}{\textbf{Speech}}  & \multicolumn{2}{c}{\textbf{Pink}} & \multirow{2}{*}{\textbf{Reverb}} & \multirow{2}{*}{\textbf{Movie}} & \multirow{2}{*}{\textbf{Babble}} \\
    {} &\textbf{Reverb} & \textbf{-5 dB} &  \textbf{0 dB} & \textbf{-5 dB} & \textbf{0 dB} & \textbf{0 dB} & \textbf{6 dB}\\
    \hline
    Baseline           & $\textbf{7.2}$ & $36.5$ & $22.5$ & $65.3$ & $44.8$ & $60.7$ & $28.4$ & $6.4$ & $80.2$ & $61.3$ \\
    \hline
    Cleanformer TasNet & $7.3$ & $17.5$  & $12.6$  & $26.3$  & $20.0$  & $29.3$ & $15.3$ & $6.3$  & $65.5$  & $50.5$\\
    Cleanformer        & $7.3$ & $13.7$ & $10.6$  & $19.8$ & $15.9$  & $19.1$ & $10.1$ & $5.8$  & $39.1$  & $31.9$\\
    + MSP E1 [this work] & $7.3$ & $13.4$  & $10.5$ & $19.2$ & $\textbf{15.5}$  & $17.9$ & $9.6$ & $5.8$ & $\textbf{33.0}$  & $\textbf{27.6}$\\
    + MSP E2 [this work] & $7.3$ & $\textbf{13.3}$  & $\textbf{10.4}$  & $\textbf{19.1}$  & $\textbf{15.5}$  & $\textbf{17.5}$ & $\textbf{9.4}$ & $\textbf{5.7}$  & $\textbf{33.0}$  & $28.5$\\
    \hline
  \end{tabular}
   \vspace{-0.15in}
\end{table*}

\subsection{Effect of Mask Scalar on ASR}\label{sec:results-alpha}
\begin{table}[tbh]
  \centering
  \caption{Performance of Cleanformer as a function of mask scalar $\alpha$; mask floor $\beta = 0.01$.}
  \label{tab:mask-scalar}
  \begin{tabular}{l|c|ccccc}
    \hline
    \multirow{2}{*}{\textbf{Condition}} & \multirow{1}{*}{\textbf{Base-}} & \multicolumn{5}{c}{\textbf{$\alpha$}} \\
    {} & {\textbf{line}} & 1e-6 & 0.25 & 0.5 & 0.75 & 1.0 \\
    \hline
    Reverb & $\textbf{7.2}$ & $7.3$ & $7.3$ & $7.3$ & $7.4$ & $7.3$ \\
    {0 dB Env.} &  $22.5$ & $22.5$ & $12.6$ & $\textbf{10.6}$ & $\textbf{10.6}$ & $11.1$ \\
    {0 dB Sp.} & $44.8$ & $44.8$ & $20.5$ & $15.9$ & $\textbf{15.8}$ & $16.7$ \\
    \hline
  \end{tabular}
   \vspace{-0.1in}
\end{table}

In Tab.~\ref{tab:mask-scalar}, we show how a fixed $\alpha$ affects ASR performance, when using Cleanformer for enhancement. Results are shown in Reverb and 0~dB SNR conditions. $\alpha=0.5$ and $0.75$ work well across conditions. When $\alpha < 0.5$, the amount of residual noise after enhancement deteriorates performance. And when $\alpha$ is closer to 1.0, the resulting speech distortion worsens WER by 4.7\% in environmental noise and 5.0\% in multi-talker conditions. Clearly, the choice of $\alpha$ significantly affects WERs. When not predicting $\alpha(\TI)$ using a mask scalar net, we conservatively set $\alpha=0.5$ for the remaining experiments.

\subsection{Speech enhancement}
Enhancement results are presented in Tab.~\ref{tab:enh}. For comparison, the table also shows results using Cleanformer TasNet, which is a waveform version of Cleanformer \cite{luo2018tasnet, panchi2022waveformaec}: It uses a learnable TasEncoder layer to convert waveforms to features, followed by a conformer encoder for mask estimation, and a TasDecoder and overlap-add operation to convert back to time-domain. During training, scale invariant SNR loss and ASR loss are used. This model is also causal, and has $\sim$1.6M parameters operating on 5 msec windows with 2.5 msec overlap, which is typical. Even though the model has fewer parameters, it requires more computation than the remaining log Mel models.

Cleanformer TasNet provides large gains over the noisy baseline; e.g., for the Rerecorded v1 set at 0~dB, it improves WER by 51.7\%. On the same set, Cleanformer outperforms Cleanformer TasNet by 34.8\%, and the baseline by 68.5\%. Enhancing directly in the feature space and controlling for speech distortions, likely help Cleanformer outperform Cleanformer TasNet. We compare two strategies for mask scalar prediction (MSP). Cleanformer + MSP E1 uses the model architecture in Fig.~\ref{fig:mask-scalar}, but does not include $\StopGrads$ operation. This already provides significant improvements, e.g.\ a 6.3\% relative reduction in WER at 0~dB on Rerecorded v1 set. The relative gains are generally larger on the mismatched, rerecorded test sets compared to simulated Librispeech test sets. Cleanformer + MSP E2, which uses $\StopGrads$ provides small gains over Cleanformer + MSP E1. Compared to Cleanformer, it improves WER by 8.4\% at 0~dB in Rerecoded v1 set. The gains are more significant in harder conditions, e.g.\ Rerecorded v2 Movie noise, where Cleanformer + MSP E2 improves over Cleanformer by 15.6\%. Overall, all enhancement frontends significantly improve performance over the baseline; using a predicted mask scalar consistently outperforms other enhancement approaches across all conditions, including the typical approach of using a hand-tuned mask scalar.

\vspace{-0.1in}
\subsection{AEC}
\begin{table}[tbh]
  \centering
  \caption{AEC performance, in terms of ASR WER.}
  \label{tab:aec}
  \begin{tabular}{l|cccc}
    \hline
    \multirow{2}{*}{\textbf{Model}} & \multicolumn{4}{c}{\textbf{SER}} \\
    {} & \multicolumn{1}{c}{\textbf{-10 dB}} & \multicolumn{1}{c}{\textbf{-5 dB}} & \multicolumn{1}{c}{\textbf{0 dB}} & \multicolumn{1}{c}{\textbf{5 dB}} \\
    \hline
    Baseline & $80.5$ & $72.7$ & $58.0$ & $36.1$ \\
    \hline
    LMel-NAEC \cite{howard2021neural} & $29.8$ & $21.8$ & $16.9$ & $14.7$ \\
    Mask-NAEC    & $26.2$ & $17.4$ & $12.5$ & $10.0$ \\
    + MSP E2 [this work] & $\textbf{24.3}$ & $\textbf{16.4}$ & $\textbf{12.0}$ & $\textbf{9.8}$ \\
    \hline
  \end{tabular}
   \vspace{-0.1in}
\end{table}

Tab.~\ref{tab:aec} shows AEC performance in terms of WER. We present results using the algorithm presented in \cite{howard2021neural} (LMel-NAEC), which estimates log Mel features directly, using regression and ASR losses. The overall architecture is identical to the proposed models, except that LMel-NAEC directly predicts the log Mel spectra as opposed to a mask. Predicting the log Mel spectra makes it challenging to limit speech distortion using post-processing techniques like the ones we use with estimated masks. Compared to LMel-NAEC, predicting the mask and using a fixed $\alpha$ and $\beta$ during inference (Mask-NAEC) already reduces the WER significantly. Note that Mask-NAEC is similar to the approach in \cite{zhang2018deep}, but it additionally introduces ASR loss into AEC training \cite{howard2021neural}. Compared to LMel-NAEC, Mask-NAEC improves WER by 12.0\% at -10~dB and 32.2\% at 5~dB. Clearly, using the post-processed mask even with a fixed mask scalar and floor significantly reduces distortion, especially at higher SNRs. Finally, using the proposed mask scalar net further improves WER by 7.2\% at -10~dB and 2.1\% at 5~dB. As with the enhancement experiments, the advantage of using a predicted mask scalar increases as the SNR goes down. Compared to LMel-NAEC, the proposed model improves WER by 18.3\% at -10~dB and 33.6\% at 5~dB.

\section{Conclusions} \label{ref:concl}
Limiting speech distortions introduced by acoustic frontends for ASR is important to get the best performance out of them. In this work, we propose using a neural net to predict exponential mask scalars in an end-to-end fashion. Our results show that predicted mask scalars reduce WER on multiple robustness tasks, with almost no increase in model size. The current work only focused on predicting a per-frame exponential scalar. Future work will explore predicting the mask floor, and more fine-grained time-frequency level prediction.

\bibliographystyle{IEEEtran}

\newpage

\end{document}